\documentclass[%
 aip,apl,
 amsmath,amssymb,
 reprint,%
]{revtex4-1}

\usepackage{graphicx}
\usepackage{dcolumn}
\usepackage{bm}

\usepackage[utf8]{inputenc}
\usepackage[T1]{fontenc}
\usepackage{mathptmx}
\usepackage{subcaption}

\begin{document}

\preprint{AIP/123-QED}

\title[KPR]{Kinetic planar resonators from strongly disordered ultra-thin MoC \\superconducting films  investigated by transmission line spectroscopy}

\author{M. Baránek}

\author{P. Neilinger}%
\altaffiliation[Also at ]{Institute of Physics, Slovak Academy of Sciences, Dúbravská cesta, Bratislava, Slovakia.}
\email{pavol.neilinger@fmph.uniba.sk}
\author{D. Manca}
\author{M. Grajcar}
\altaffiliation[Also at ]{Institute of Physics, Slovak Academy of Sciences, Dúbravská cesta, Bratislava, Slovakia.}
\affiliation{ 
Department of Experimental Physics, Comenius University, SK-84248 Bratislava, Slovakia}%

\date{\today}

\begin{abstract}
The non-contact broadband transmission line flip-chip spectroscopy technique is utilized to probe resonances of mm-sized square kinetic planar resonators made from strongly disordered molybdenum carbide films, in the GHz frequency range. The temperature dependence of the resonances was analyzed by the complex conductivity of disordered superconductor, as proposed in Ref.~\onlinecite{Zemlicka15}, which involves the Dynes superconducting density of states. The obtained Dynes broadening parameters relate reasonably to the ones estimated from scanning tunneling spectroscopy measurements. The eigenmodes of the kinetic planar 2D resonator were visualized by EM model in Sonnet software. The proper understanding of the nature of these resonances can help to eliminate them, or utilize them e.g. as filters. 

\end{abstract}

\maketitle

Thin disordered superconducting films find a variety of applications in advanced technologies, such as superconducting single-photon detectors,\cite{Goltsman01} parametric amplifiers,\cite{Tholen09, Eom12} superinductors,\cite{Manucharyan09,Niepce19} and superconducting quantum bits.\cite{Astafiev12} 
These devices make use of the distinctively high sheet resistance R$_s$ of these films and thus the high kinetic sheet inductance, which is approximately related to R$_s$ and the superconducting energy gap $\Delta$ by $L_k=\hbar R_s/\pi \Delta$ at temperatures well below the superconducting critical temperature T$_c$. These devices typically operate in the GHz frequency range, and the knowledge of their complex conductivity is of crucial importance for the device design and properties. Moreover, the electrodynamic response of these films provides important insight into the fundamental topics of superconductivity, such as the superconductor-insulator quantum phase transition \cite{Gantmakher10} and the Berezinskii-Kosterlitz-Thouless transition.\cite{Beasley79} 
To probe the complex conductivity of thin films, either as surface impedance or through penetration depth measurements, several microwave spectroscopy techniques are exploited. These are either narrowband techniques, such as cavity resonators,\cite{Krupka93, Zhai00} the parallel plate resonator technique (which is commonly employed to study high-T$_c$ superconductive films), planar transmission line resonators,\cite{Scheffler12} or broadband spectroscopy techniques like the transmission line or Corbino spectroscopy.\cite{Corbino11, Booth94, Scheffler05, Kitano08} The complex conductivity of superconductors is often studied in the vicinity of the superconducting transition.\cite{Ganguly15, Nabeshima18} In several cases, resonances in the spectra were reported for metallic and superconducting films on dielectric substrates. \cite{Kitano04, Scheffler05, Kitano08, Scheffler10, Scheffler10epj, Steinberg10, Geiger12, Neilinger20} In general, the presence of these resonances is undesirable as they can strongly limit the upper frequency range of the spectroscopy. The origin of these resonances in thin metallic films on a dielectric sample was unveiled by simulations and experiments as substrate and cavity resonances\cite{Felger2013} and analyzed in detail by neglecting the metallic films.  

In this letter, we report on resonances present in the transmission spectra of disordered superconducting films of molybdenum carbide (MoC) measured by broadband transmission line spectroscopy in temperature range from T$_c$ down to $\approx$ 500 mK and frequency range from 1~GHz to 16~GHz. The measurements were carried out in flip-chip geometry. The 5~nm thin films with different R$_s$ were sputtered on sapphire substrate and placed on top of the transmission line facing the center conductor. The present resonances are explained by referring them to eigenmodes of a kinetic planar 2D resonator. Low resonance frequencies of the resonator are the result of the high kinetic inductance of the strongly disordered MoC films. 
Moreover, in contrast to the aforementioned works, we utilize these resonances to determine the Dynes broadening parameter $\Gamma$ of these films, which is usually determined by scanning tunneling spectroscopy (STS) at temperatures well below T$_c$. The superconducting density of states (SDOS) of disordered superconductors is characterized by the smearing of coherence peaks and the presence of in-gap states. Their STS spectra are generally analyzed in terms of the Dynes  SDOS\cite{Dynes78} $N(E) = Re\{(E+i\Gamma)/[(E+i\Gamma)^2-\Delta^2]^{1/2}\}$. 
A thorough study of the STS spectra of highly disordered MoC films is presented in Ref.~\onlinecite{Szabo16}. As a result of disorder, the temperature and frequency dependent complex conductivity  $\sigma=\sigma_1-i\sigma_2$ deviates from the well known Mattis-Bardeen  conductivity,\cite{Driessen12, Coumou13, Zemlicka15} in contrast to clean BCS superconductors. These deviations were directly related to the complex conductivity of disordered superconductors \cite{Zemlicka15} utilizing Nam's theory. \cite{Nam67} This approach was furthered to the optical conductivity of disordered superconductors.\cite{Herman17} 

A series of four MoC films was sputtered by means of reactive magnetron deposition from a molybdenum target, in argon-acetylene atmosphere on top of 5$\times$5~mm c-cut sapphire substrate. The R$_s$ of the MoC film was subsequently increased by measuring the carbon content, controlled via the argon-acetylene flow in the chamber during the deposition.\cite{Neilinger20} The increased disorder results in increased R$_s$ and the suppression of T$_c$. Eventually, at critical R$_s$,  these films undergo a superconductor-insulator transition.\cite{Lee90} The properties of the samples are listed in the Table~\ref{table:parameters}. R$_s$ is measured by Van der Pauw method at room temperature. As the transition of highly disordered superconductors, measured by the DC transport, is broadened, the maximum of the resistivity temperature derivative T$_c^{DC}$  is listed.

\begin{table}
\caption{\label{table:RT}Sample parameters: R$_s$ is  sheet resistance at room temperature, T$_c^{DC}$ is superconducting critical temperature from transport measurement, T$_c^{spec}$, $\Gamma^{spec}$, and $\Gamma^{STS}$ are values estimated from spectroscopic measurement in the GHz frequency range and from STS measurements, respectively.}
\begin{ruledtabular}
\begin{tabular}{cccccc}
Sample & R$_s$ ($\Omega$) & T$_c^{DC}$ (K) & T$_c^{spec}$ (K)  & $\Gamma^{spec}/\Delta_0$ & $\Gamma^{STS}/\Delta_0$\\ \hline
A & 120 & 7.89  & -     & -   & 0  \\
B & 212 & 7.04  &7.02   & 0   & 0.03\\
C & 565 & 4.85  &5.11   & 0.2 & 0.20\\
D & 974 & 2.44  &2.52   & 0.5 & 0.44\\

\end{tabular}
\end{ruledtabular}
\label{table:parameters}
\end{table}

\begin{figure}[h]

\begin{subfigure}{0.5\columnwidth}

  \includegraphics[width=\columnwidth]{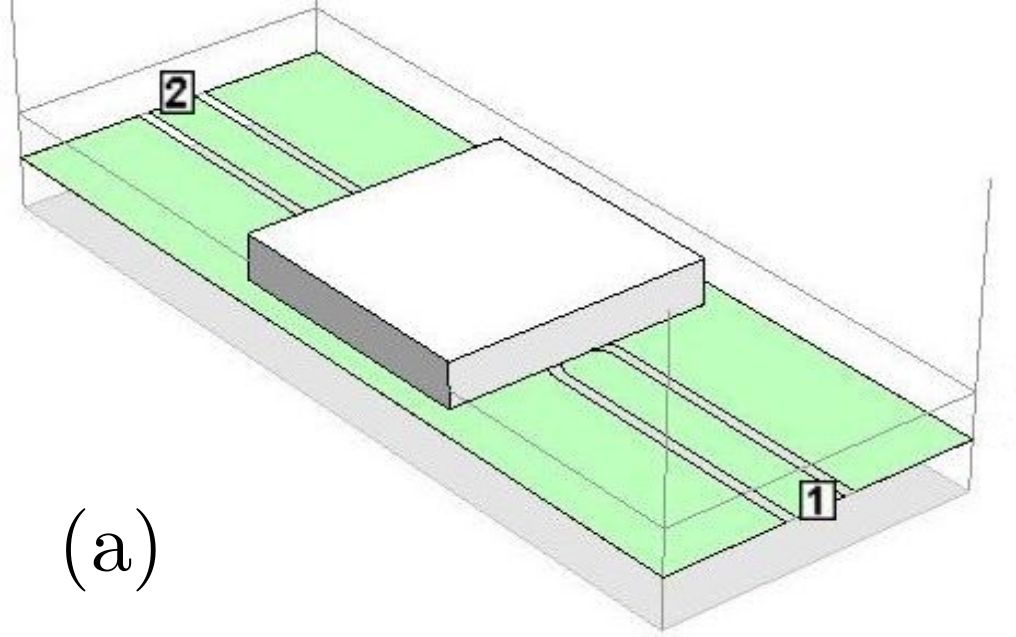}%
\end{subfigure}%
\begin{subfigure}{0.5\columnwidth}
  \includegraphics[width=\columnwidth]{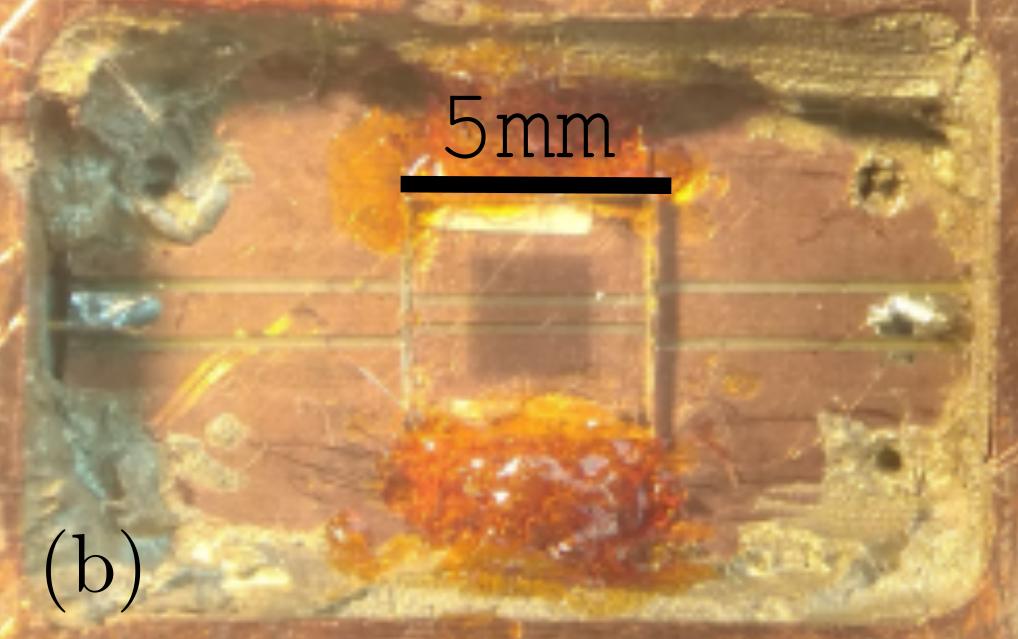}
\end{subfigure}
\caption{\label{fig:exp_sample}(Color online) (a) The 3D model of the coplanar waveguide created in SONNET software. The superconducting film is at the bottom of the sapphire substrate. (b) Photo of a rectangular film sample suspended above the transmission line.}
\end{figure}

Next, the samples were measured by broadband non-contact flip-chip transmission line spectroscopic technique, where the sample is placed on top of the transmission line facing the center conductor. The model and the experimental transmission line are shown in Fig.~\ref{fig:exp_sample}.

\begin{figure}
\begin{subfigure}{0.5\columnwidth}
  \includegraphics[width=\columnwidth]{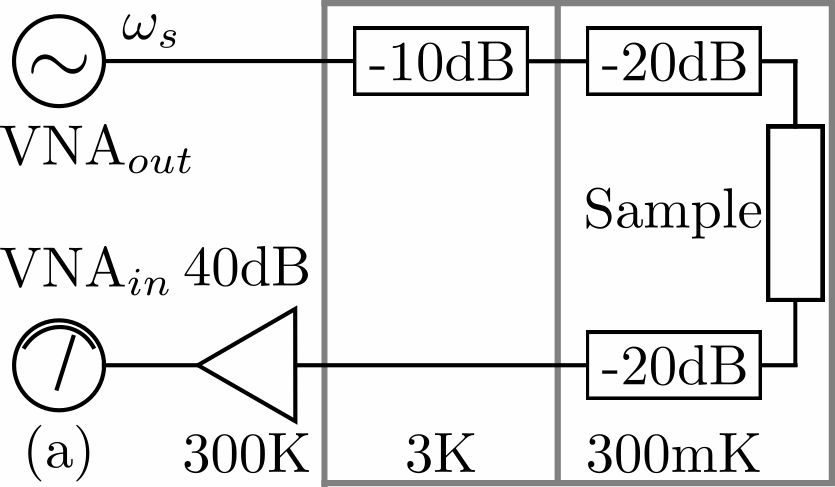}%

\end{subfigure}%
\begin{subfigure}{0.5\columnwidth}
  \includegraphics[width=\columnwidth]{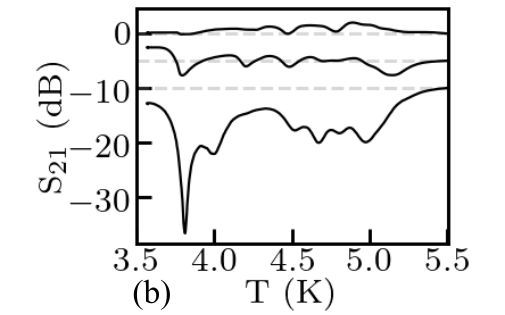}
\end{subfigure}

\caption{\label{fig:2}(a) The scheme of the experimental set-up for transmission measurement. (b) Temperature dependence of the normalized transmission spectra from top to bottom, at 3, 6 and 9 GHz of sample C. The curves are offset by -5dB.}

\end{figure}

The coplanar waveguide was fabricated on a Rogers RO4003 PCB. The tapering in the middle of the transmission line ensures impedance matching upon placing a bare 5$\times$5~mm$^2$ sapphire substrate with thickness $460$~$\mu$m above the tapering. The bandwidth of the transmission line is up to 18~GHz. The transmission line was fixed to a copper box by silver epoxy and soldered to a SMA Through Hole connector. To ensure electrical insulation, the sample was placed on a thin cigarette paper and fixed by Ge-varnish.
The temperature dependence measurements were carried out in a $^3$He refrigerator and the transmission spectrum was measured by a vector network analyzer, see Fig~\ref{fig:2}a. 
Whereas the transmission spectra of sample A with the lowest R$_s$ show the expected step-like increase in the transmission below T$_c$, the temperature dependence of the spectra for sample B with R$_s$=212$\Omega$ is slightly disturbed by unexpected resonances close to T$_c$. The frequency of these resonances increases with decreasing temperature, indicating their dependence on the superconducting properties of the film. Further, these resonances gain in strength and shift to lower frequencies with the increase of R$_s$, following the $L_k=\hbar R_s/\pi \Delta$ dependence, suggesting their origin in the resonances of the superconducting film.

\begin{figure}
\includegraphics[width=\columnwidth]{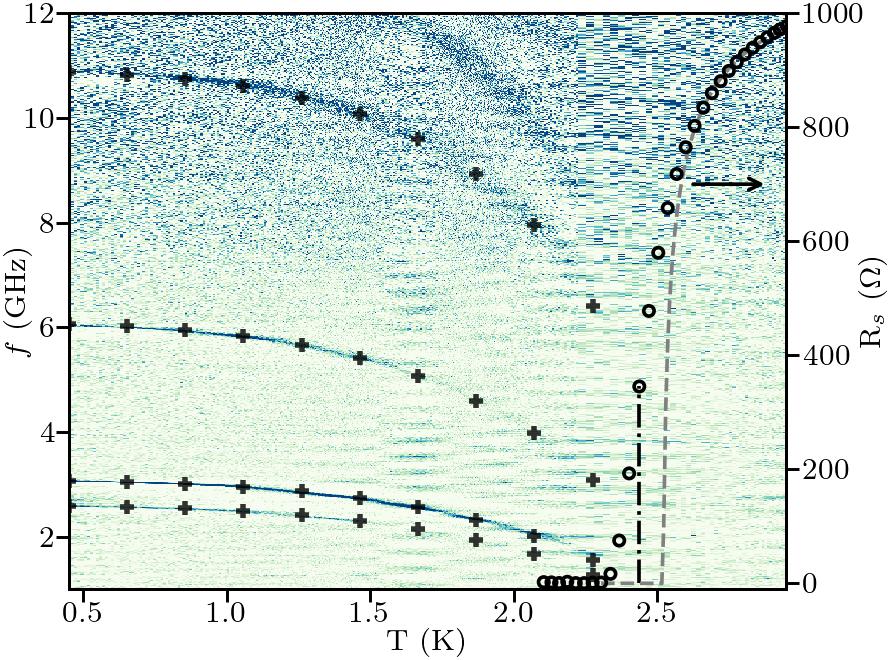}
\caption{\label{fig:results} (Color online)
Normalized transmission spectra of the sample D, showing temperature-dependent resonances; fit of the resonances ($+$); temperature dependence of $R_s$ ($\circ$); The $T_c^{DC}$ from DC measurement is denoted be dash-dotted line and the grey dashed line is the fit of DC transition following \cite{Benfatto09}. For presentation, Savitzky–Golay filter was used to suppress the background.}
\end{figure}

To further prove this assumption, samples C and D were shaped by optical lithography and dry etching process onto a square with 2.25~mm sidelength. Indeed, this resulted in the increase of the frequencies co-responding to the geometrical factor and suppressed T$_c$. Furthermore, the well-defined shape of the samples increased the quality of the resonances, showing the importance of the edges, typical for resonators. The temperature dependence of the transmission spectrum for sample D is shown in Fig. \ref{fig:results}. 

The spectrum is normalized to the normal state transmission. As it is visible, several resonances emerge from $\omega \xrightarrow{} 0$ at T$_c$ and with decreasing temperature, they shift toward higher frequencies. To analyze the resonances, we model them as simple capacitively-coupled resonance modes. Their temperature dependent frequency can be approximated, assuming $R \xrightarrow{} 0$, with the LC circuit resonance equation\cite{Goeppl08} ($S_{21}(\omega_n) = 0$):

\begin{equation}
\omega_n(T) \approx \frac{1}{\sqrt{((C_r+C_{c})L_{r})_n+ (C_r+C_{c})_n L_k(\omega_n,T)}},
\label{eq:1/LC}
\end{equation}
where C$_r$ and L$_r$ are geometric parameters defining the resonance mode. Assuming these parameters are constant for the n$^{th}$ resonance mode in the measured narrow temperature range, the temperature dependence can be expressed by the simple equation:
\begin{equation}
\omega_n(T) \approx \frac{1}{\sqrt{A_n + B_n L_k(\omega_n,T)}},
\label{eq:1/ABLC}
\end{equation}
where $A_n,B_n$ are fitting parameters, specific for each resonance mode. 

The kinetic sheet inductance of a thin superconducting film with thickness much smaller than the London penetration depth is given as\cite{Gao08} $L_k(\omega,T) = \sigma_2 (\omega_n,T)/\omega\sigma^2 (\omega_n,T)$. The complex conductivity of the films with finite $\Gamma$ are numerically calculated\cite{Zemlicka15} for a set of $\Gamma$, and the full superconducting gap $\Delta_0 =\Delta(T=0) $ parameters and according to Eq.~\ref{eq:1/ABLC}, the temperature dependence of the resonances in the transmission spectra are fitted, resulting in the corresponding values of $\Gamma$ and $\Delta_0$. For MoC films, the relation $\Delta_0 = 1.84T_c$ is used\cite{Szabo16} and the temperature dependence of $\Delta(T)$ follows the BCS relation.

The fitted curves for sample D are shown in Fig.~\ref{fig:results}. The estimated values of T$_c$ and $\Gamma$ resulting in the best fit, together with the expected values of $\Gamma$ estimated from STS\cite{Szabo16} spectra are listed in Table~\ref{table:parameters}. The values of $\Gamma$ obtained  from spectroscopy correspond reasonably to the ones obtained from STS. The estimated $T_c$ is slightly above the values estimated from DC transport measurement, but in the broadened superconducting transition.
The complex conductivity of Dynes superconductors reproduces the experimentally observed temperature dependence in the low-temperature limit, where MB conductivity model ($\Gamma \xrightarrow{} 0$) is already saturated. As a result, the MB conductivity fit underestimates the T$_c$ and fits the experimental dependence with several times the error of the model with finite $\Gamma$.

\begin{figure}
\includegraphics[width=\columnwidth]{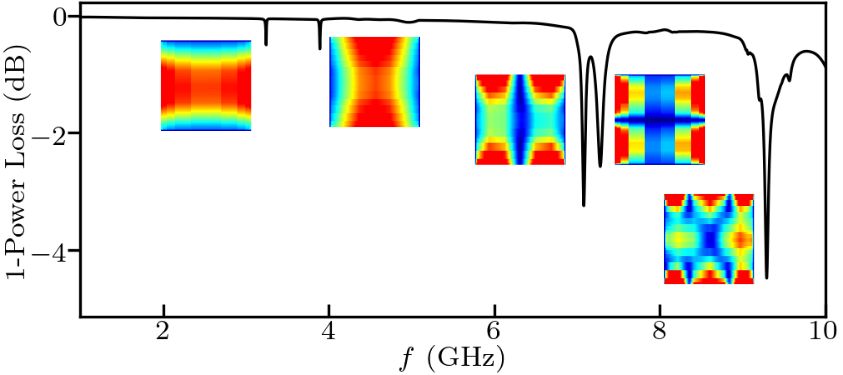}
\caption{\label{fig:spectrum}(Color online) Simulated power loss in the trasmission line and the corresponding eigenmodes of the planar resonator. The relative amplitude of the  $j_y,j_x,j_x,j_y,j_x$ current densities in superconducting layer are color coded.}
\end{figure}

To further investigate the nature of the resonances, the measurement was modeled in EM software Sonnet (Fig.~\ref{fig:exp_sample}a). The calculated complex conductivity of the film was inserted into the software. The model reproduced the measurement qualitatively, showing resonant absorption dips in the transmission spectra. The resonances below 10~GHz are governed by the thin film properties. Above 10~GHz, the metal box resonances and the dielectric resonances of the substrate are present, too. The latter resonances were recognized and analyzed in Ref.~\onlinecite{Felger2013} for experiment with thin metallic films on dielectric substrates. The kinetic inductance and geometric inductance of the metallic films are negligible in comparison to the disordered superconductors, and their resonances would be present at much higher frequencies. The high L$_k$ of MoC films results in planar resonances present at frequencies as low as 3~GHz, despite the small dimensions of the resonator. The simulated power loss and the corresponding resonances are shown in Fig.~\ref{fig:spectrum}. The dominant current component in the superconducting film for the first 5 resonant modes illustrates the nature of the resonances (Fig. \ref{fig:spectrum}).
The frequencies of the resonances do not fit the experiment perfectly, partly due to the lack of experimental control in sample position limiting the accuracy of the model, and partly due to the difficulty of the simulations. However, they exhibit the same temperature dependence as in the experiment.

To conclude, we utilized a broadband transmission line flip-chip spectroscopy technique to study the superconducting transition of thin, strongly disordered MoC films sputtered on sapphire substrates. The spectra contained unexpected resonances, which were identified as planar resonances of the superconducting film. These resonances have low resonance frequency due to the high L$_k$ of disordered superconductors and they correspond to the eigenmodes of the planar 2D resonator. In contrast to previous works\cite{Felger2013}, the dynamics of the resonances is given by the complex conductivity of the disordered films and dimensions of the films. This is further supported by the EM model in Sonnet software. The resonances were observed on rectangular $\approx$ 5$\times$5~mm$^2$ films set by the substrate solely, and on optically lithographed 2.25$\times$2.25~mm$^2$ squares with different R$_s$. 
The temperature dependence of the resonances was fitted by numerically calculated complex conductivity of disordered superconductor\cite{Zemlicka15} with finite $\Gamma$. The obtained $\Gamma$ parameters relate reasonably to the ones expected from STS measurements, showing that this theory could be utilized to describe the electrodynamic response of highly disordered superconductors, and vice versa: this non-contact method could be used to estimate the $\Gamma$ and T$_c$ of highly disordered films, avoiding the nontrivial calibration procedure required in similar spectroscopic techniques\cite{Zinsser19}, or the requirement of structuring special resonators on the films\cite{Wendel20}. Moreover, the understanding of the nature of these resonances can help eliminate them in the required bandwidth by smart design, as for example in Ref.~\onlinecite{Naji19}, or to utilize them as filters in sensors and QED circuitry design. Their possible advantage could be a higher power handling capability than their quasi 1D on-chip microwave counterparts\cite{Collard20} and their disjunction from the circuitry design.  Finally, the bare awareness of the existence of these resonances in highly disordered films is important, as they can affect spectroscopic experiments, where strongly disordered films are studied and delicate deviations in the complex conductivity close to T$_c$ are analyzed \cite{Ganguly15}.

\begin{acknowledgments}
This work was supported by the Slovak Research and Development Agency under the contract APVV-16-0372,  APVV-18-0358 and by the QuantERA grant SiUCs, by SAS-MTVS.
\end{acknowledgments}

\nocite{*}
\bibliography{aipsamp}

\end{document}